\documentclass[aps,prd,preprint,showpacs,floatfix,nofootinbib,superscriptaddress]{revtex4-1}

\usepackage{dcolumn}

\usepackage{amsmath, amsfonts, amssymb, mathrsfs, times, float, color, bm}
\usepackage{extarrows}
\usepackage{graphicx}
\usepackage{graphicx,epstopdf}
\usepackage{bm}

\usepackage{exscale}
\usepackage{relsize}

\begin{document}
\title{Exact solution for Schwarzschild black hole in radiation gauge}

\author{Wenbin Lin}
\email{wl@swjtu.edu.cn}
\affiliation{School of Physical Science and Technology, Southwest Jiaotong University, Chengdu 610031, China}

\author{Chunhua Jiang}
\affiliation{School of Mathematics and Physics, University of South China, Hengyan 421001, China}
\begin{abstract}
Recently Chen and Zhu propose a true radiation gauge for gravity [Phys. Rev. D 83, 061501(R) (2011)]. This work presents a general solution for the metric of Schwarzschild black hole in this radiation gauge.
\begin{description}
\item[PACS numbers] 04.20.-q, 04.20.Cv, 04.20.Jb 
\end{description}
\end{abstract}

\maketitle
\section{Introduction}

The space-time metric is determined by the matter energy-momentum tensor of gravitational source via Einstein field equations
\begin{equation}
R_{\mu\nu}-\frac{1}{2}g_{\mu\nu}R=-8\pi G T_{\mu\nu}~,\label{EE}
\end{equation}
where $g_{\mu\nu}$ is the metric tensor, $R_{\mu\nu}$ and $R$ are Ricci tensor and scalar, and $T_{\mu\nu}$ denotes the matter energy-momentum tensor. The light speed $c$ in vacuum is set as $1$. Here and in the following, Greek indices run from $0$ to $3$ and Latin indices run from $1$ to $3$.

Due to Bianchi equalities, Einstein field equations only have six independent equations, while, a metric tensor has ten components. Therefore, we need four additional conditions to fix the solution when we solve Einstein field equations. This is also called as gauge fixing. 

Although it is generally believed that the gauge can be chosen arbitrarily, Fock believes that the solution to Einstein field equations has physical significance only under the harmonic-coordinate conditions~\cite{Fock1957},
\begin{equation}
g^{\mu\nu}\Gamma^{\lambda}_{\mu\nu}=0~,\label{HG}
\end{equation}
where $\Gamma^{\lambda}_{\mu\nu}$ denotes the affine connection. The harmonic-coordinate conditions have been employed in obtaining the post-Newtonian solution to Einstein field equations and dealing with the calculation of the gravitational radiation. The harmonic-coordinate conditions play a similar role as does the Lorentz gauge $\partial_{\mu}A^{\mu}=0$ with $A^{\mu}$ being the vector potential in electrodynamics, and they does not fix the gauge completely and are not suficient to specify the two physical polarizations of the gravitational wave. Recently, Chen and Zhu pioneer a true radiation gauge for general relativity~\cite{ChenZhu2011},
\begin{equation}
g^{ij}\Gamma^{\lambda}_{ij}=0~,\label{TG}
\end{equation}
which is similar to the Coulomb gauge $\partial_{i}A^{i}=0$ in electrodynamics. This gauge can remove all nonphysical degrees of freedom of the gravitational field in the weak-field limit, and thus the energy of gravitational wave can be calculated directly. Therefore, this gauge may be viewed as an improvement over the harmonic gauge for dealing with the gravitational wave radiation, and would make the canonical structure and quantization of gravity more illuminating~\cite{ChenZhu2011}. In their work, the authors focus on the weak-field regime and present the radiation-gauge solution for the metric of Schwarzschild black hole to the first order. It is interesting and important to study the general solution for Schwarzschild black hole in this radiation gauge, and this is the aim of this work.

\section{Derivation of exact solution for Schwarzschild black hole in the radiation gauge}
We start with the standard form of Schwarzschild metric~\cite{Schwarzschild}
\begin{equation}
ds^2= - \left(1-\frac{2 G M}{R}\right)dt^2 + \left(1-\frac{2 G M}{R}\right)^{-1}dR^2 + R^2 \left(d\theta^2 + \theta^2 d\varphi^2 \right)~,\label{SM}
\end{equation}
where $M$ is the mass of black hole. In order to find the solution satisfying the radiation gauge Eq.~(\ref{TG}), we introduce a new radius variable $r$, and let $R$ be a function of $r$. Re-writing Eq.~(\ref{SM}) in terms of $r$, we have
\begin{equation}
ds^2= - \left(1-\frac{2 G M}{R}\right)dt^2 + \left(1-\frac{2 G M}{R}\right)^{-1} R'^2 dr^2 + R^2\left(d\theta^2 + \sin^2\theta d\varphi^2 \right)~,\label{SM1}
\end{equation}
where the prime denotes the derivative with respect to $r$.

Similar to the construction of the harmonic coordinates for Schwarzschild metric~\cite{Weinberg1972}, we can construct the coordinates in the radiation gauge as follows:
\begin{equation}
x^1 = r \sin\theta\cos\varphi~, ~~~~ x^2 = r \sin\theta\sin\varphi~, ~~~~ x^3 = r \cos\theta~, ~~~~x^0 = t~.\label{SM3}
\end{equation}
Then Eq.~(\ref{SM1}) becomes
\begin{equation}
ds^2= - \left(1-\frac{2 G M}{R}\right)dt^2 + \left[ \left(1-\frac{2 G M}{R}\right)^{-1} \frac{R'^2}{r^2} - \frac{R^2}{r^4} \right] (\bm{x} \!\cdot\! d\bm{x})^2 + \frac{R^2}{r^2}d\bm{x}^2 ~.\label{SM4}
\end{equation}
The components of the metric can be written down explicitly,
\begin{eqnarray}
&& g_{00} = - \left(1-\frac{2 G M}{R}\right)~,\label{metric00} \\
&& g_{0i} = 0~,   \\
&& g_{ij} =\frac{ R^2}{r^2}\delta _{ij} + \left[R'^2 \left(1-\frac{2 G M}{R}\right)^{-1}-\frac{R^2}{r^2}\right]\frac{x^i x^j}{r^2} ~.\label{metricij}
\end{eqnarray}

In order to derive the general solution, we also need the inverse of metric, which can be obtained from Eqs.~(\ref{metric00})-(\ref{metricij}) as follows:
\begin{eqnarray}
&& g^{00} = - \left(1-\frac{2 G M}{R}\right)^{-1}~, \label{metric^00} \\
&& g^{0i} = 0~, \\
&& g^{ij} = \frac{r^2}{R^2}\delta _{ij}- \left[1-\frac{R^2}{R'^2r^2} \left(1-\frac{2 G M}{R}\right) \right]\frac{x^i x^j}{R^2}~.\label{metric^ij}
\end{eqnarray}

Substituting Eqs.~(\ref{metric00})-(\ref{metric^ij}) into Eq.~(\ref{TG}), after tedious but straightforward calculation, we find the 0-component of Eq.~(\ref{TG}) is automatically satisfied, and the $l$-commponent can be written as
\begin{equation}
\left[ \left( 1-\frac{2GM}{R} \right) \frac{R''}{R'r} +\left( \frac{2R'}{R} - \frac{2}{r} + \frac{3GM}{Rr} \right)\frac{R'}{R} \right]\frac{2{x}^l}{R'^2} =0
\end{equation}
Therefore, we have
\begin{equation}
\left( 1-\frac{2GM}{R} \right) \frac{R''}{R'r} +\left( \frac{2R'}{R} - \frac{2}{r} + \frac{3GM}{Rr} \right)\frac{R'}{R} =0~. \label{TGRequired}
\end{equation}

One simple solution to Eq.~(\ref{TGRequired}) can be written as
\begin{equation}
R= \alpha r + \frac{3}{2} G M~,\label{solc}
\end{equation}
with $\alpha$ being a constant. Substituting this equation into Eq.~(\ref{SM4}), we have 
%
\begin{equation}
ds^2 \! = \! -\left( 1\! - \!\frac{2 G M}{\alpha r \!+\! \frac{3}{2}G M }\right)  dt^2 +
\left(\alpha\!+\!\frac{3}{2} \frac{G M}{r} \right)^2 \! d\bm{x}^2
 +\left[\frac{-GM (\alpha r \!-\! \frac{3}{4} GM  ) (\alpha r \!+\! \frac{3}{2} GM)}{ r^2 (\alpha r \!-\! \frac{1}{2}GM  )}\right] \frac{(\bm{x} \!\cdot\! d\bm{x})^2}{r^2}~.\label{Metric_RGalpha}
\end{equation}

The constant $\alpha$ can be fixed when we consider the weak-field limit of general relativity, which should be consistent with Newtonian theory. In the limit of $r\rightarrow \infty$, the time-time component of Einstein field equation gives~\cite{Weinberg1972}
\begin{equation}
\nabla^2 g_{00} = -8\pi G T^{00}~.\label{new1}
\end{equation}
Taking the weak-field limit of Eq.~\eqref{Metric_RGalpha}, and plugging the time-time metric component into Eq.~\eqref{new1}, we have
\begin{equation}
\nabla^2 \left(\frac{2GM}{\alpha r}\right) =-8\pi G M\delta(r)~,
\end{equation}
and we can obtain $\alpha=1$.

Therefore, a general solution for Schwarzschild black hole in the radiation gauge can be written as follow:
\begin{equation}
ds^2 \! = \!- \frac{1\!-\!\frac{1}{2} G M/r}{1\!+\! \frac{3}{2} G M/r} dt^2 \!+\! \left(1\!+\!\frac{3}{2} \frac{G M}{r}\right)^{\!2} \! d\bm{x}^2 \!+\! \left[\frac{1\!+\!\frac{3}{2} G M/r}{1\!-\!\frac{1}{2}G M/r} \!-\! \left(\!1\!+\!\frac{3}{2}\frac{G M}{r}\right)^{\!2} \right]\!\frac{(\bm{x} \!\cdot\! d\bm{x})^2}{r^2}~.\label{Metric_RG}
\end{equation}
where $r^2=\bm{x}^2$~. It is easy to check that this solution reduces to the first-order result given in the work~\cite{ChenZhu2011} in the weak-field limit. 


\section{Summary}
In this work we have obtained the exact metric of Schwarzschild black hole in the radiation gauge, and this solution may be useful in studying gravitational physics near the black hole. 


\section*{ACKNOWLEDGEMENT}
This work was supported in part by the National Natural Science Foundation of China (Grant Nos. 11547311 and 11647314).


\end{document}